\documentclass[letterpaper,12pt,pra]{revtex4}
\usepackage{epsfig,amsfonts,amsmath}
\usepackage{graphics,times,hyperref,fullpage}
\usepackage{setspace} 
\usepackage[margin=2.6cm]{geometry}
\newcommand{\be}{\begin{equation}}
\newcommand{\ee}{\end{equation}}
\newcommand{\ba}{\begin{eqnarray}}
\newcommand{\ea}{\end{eqnarray}}

\def\bea{\begin{eqnarray}}
\def\eea{\end{eqnarray}}
\def\ben{\begin{eqnarray*}}
\def\een{\end{eqnarray*}}

\def\>{\rangle}
\def\<{\langle}

\newcommand{\eq}[1]{Eq.~(\ref{eq:#1})}
\newcommand{\fig}[1]{Fig.~\ref{fig:#1}}
\newcommand{\secref}[1]{Sec.~\ref{sec:#1}}
\newcommand{\tab}[1]{Table~\ref{tab:#1}}

\begin{document}

\title{Ideal Multipole Ion Traps from Planar Ring Electrodes}
\author{Robert J. Clark}

\affiliation{Department of Physics, The Citadel, Charleston, SC 29409}
\date{\today}

\begin{abstract}
We present designs for multipole ion traps based on a set of planar, annular, concentric electrodes which require only rf potentials to confine ions. We illustrate the desirable properties of the traps by considering a few simple cases of confined ions. We predict that mm-scale surface traps may have trap depths as high as tens of electron volts, or micromotion amplitudes in a 2-D ion crystal as low as  tens of nanometers, when parameters of a magnitude common in the field are chosen. Several example traps are studied, and the scaling of those properties with voltage, frequency, and trap scale, for small numbers of ions, is derived. In addition, ions with very high charge-to-mass ratios may be confined in the trap, and species of very different charge-to-mass ratios may be simultaneously confined.  Applications of these traps include quantum information science, frequency metrology, and cold ion-atom collisions. 
\end{abstract} 

\maketitle

\section{Introduction}\label{sec:motiv}

Ion traps are ubiquitous tools in physics and chemistry, with applications including quantum information science \cite{Haeffner_habil:08}, precision measurement \cite{Schmidt:05,Rosenband:08}, and mass spectrometry \cite{Douglas:05}. The two main classes of ion trap are Paul traps, which use an oscillating electric field, usually combined with a static electric field, and Penning traps, which combine a static magnetic field with a static electric field. In many  applications, such as quantum information science, the ions are laser-cooled, which can cause them to condense into an ordered crystal. Even at low temperature, however, the ions in both types of trap suffer an unwanted excess motion: rf-driven micromotion, in the case of Paul traps, and rotation of the crystal due to the crossed electric and magnetic fields in Penning traps. In linear Paul traps, the ions form a one-dimensional chain, and micromotion along the axis of the chain is minimal, but this is not so for crystals of two and higher dimensions. So far, there has been no method of preparing a 2D or 3D crystal of motionless charged particles in a single trapping region. 

One approach to building a trap in which ions are largely free from excess motion is a so-called \emph{multipole} ion trap. The most common variety of such, the linear multipole trap, is made of several parallel rods, the voltage on each being either  180$^{\circ}$ out of phase with its neighbor or grounded. Although laser-cooled crystals have been observed in such linear multipole traps \cite{Okada:07}, one encounters the unwanted effect of the total potential minimum being displaced from the minimum of the rf pseudopotential by the static axial confining field \cite{Champenois:10}. This leads to greater micromotion than would otherwise exist. Two particular deleterious effects of this micromotion are errors related to the Doppler shift in atomic clocks based on trapped ions \cite{Champenois:10}, and a fundamental upper limit to the temperatures of ion-atom mixtures \cite{Cetina:12}. A recent report shows that micromotion-free parallel ion strings can be formed with additional rf potentials \cite{Marciante:11}. However, we are interested in the question of whether it is possible to create a multipole trap potential with a minimum on the axis of symmetry, and in which all ions occupy a single trapping volume. We refer to this situation as an \emph{ideal} multipole ion trap. 

Our solution to this problem is inspired by novel ion trap designs based on electrodes that lie in a single plane \cite{Chiaverini:05,Seidelin:06,Pearson:06}. These \emph{surface-electrode} or \emph{planar} traps can be microfabricated using standard optical lithography. The advantages of miniaturization include higher interaction rates, integration of control fields into the trap structure \cite{Chiaverini:08,Ospelkaus:08,Mintert:01}, and the possibility of moving ions between memory and processing zones \cite{Kielpinski:02}. A \emph{layered} planar multipole trap has been developed \cite{Debatin:08};  however, its linear geometry implies that axial confinement is still necessary, both leading to rf-driven motion and impeding optical access. We therefore ask the question of whether a surface-electrode multipole trap (SEMT) can be built. 

The remainder of this article is organized as follows. In \secref{approach}, we  outline the general mathematical approach to the problem. In \secref{solving}, we  present solutions obtained in a number of ways for SEMT's. \secref{properties} is devoted to understanding the properties of the traps, in particular relating experimentally adjustable parameters such as the drive frequencies and voltages to properties such as the trap depth and micromotion amplitude for a particular ion. We conclude with \secref{discuss}, which contains a discussion of the possible applications of these traps. 

\section{General Approach}\label{sec:approach}

\begin{figure}

\begin{center}
\includegraphics[width=.5\textwidth]{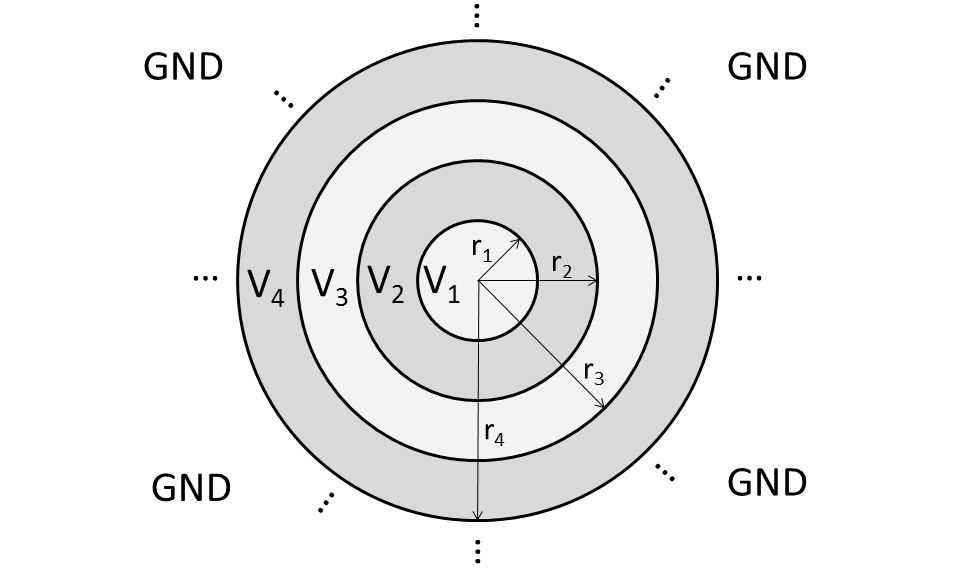}
\end{center}
\caption{ Our model trap consists of a set of planar, annular electrodes. The cylindrical coordinate $r$ is measured from the center of this structure, and the coordinate $z$ measures distance above this structure. The electrode widths are arbitrary and are specified here using $r_1$, $r_2$, etc., as shown, which is possible in the gapless approximation which we use. The voltages on each electrode $V_1$, $V_2$, ... may be set to any value; the ``ellipses'' indicate that an arbitrary number of rings may be used. The ring electrodes are assumed to be surrounded by an infinite grounded ($V = 0$) plane. \label{fig:new_layout}}
\end{figure}

This work deals exclusively with traps based on concentric rings. This choice is based on the fact that a ring geometry allows one to trap ions in three dimensions using only rf (and no nonzero dc) potentials \cite{THKim:10}. Our model trap (\fig{new_layout}) consists of a set of flat conducting rings to which either an rf voltage or zero voltage is applied. We assume that the distance between electrodes is zero and that the outer electrode is surrounded by a plane of infinite extent. Much work has been published on analytical solutions for the electric potential of surface-electrode ion traps \cite{House:08,Wesenberg:08,Schmied:10,THKim:10,Stahl:05}; here, we find it convenient to use the form of the solutions presented in Ref.~\cite{THKim:10}. 

We first review the general solution for our trap geometry. Employing cylindrical coordinates $z$ and $r$, where the origin for $z$ is on the trap surface and for $r$ is on the axis of symmetry, the electric potential is written 

\begin{equation}
\Phi(z,r) = \int_0^{\infty} J_0(kr) e^{-kz} A_0(k) dk  
\label{eq:phi} 
\end{equation} 

\noindent where $J_0$ is the Bessel function of zeroth order and $A_0(k) = \sum_{i=1}^N A_i(k)$. The $A_i$ are given by 

\begin{equation}
A_i(k) = V_i \left ( r_{i+1} J_1(kr_{i+1}) - r_i J_1(kr_i) \right ) ,
\end{equation} 

\noindent where $V_i$ is the amplitude of the rf voltage applied to the $i^{th}$ electrode, $J_1$ is the Bessel function of first order, and the $r_i$ are the radii describing the trap geometry (as shown in \fig{new_layout}). In our calculations, we make the pseudopotential approximation, in which the time-averaged pseudopotential is given by 

\begin{equation}
\Psi(z,r) = \frac{Q^2}{4 m \Omega^2} |\vec{E}(z,r)|^2 , 
\label{eq:pseudopot} 
\end{equation}

\noindent where $Q$ is the ion's charge, $m$ is the ion's mass, $\Omega$ is the frequency of the rf potentials applied to the electrodes, and $\vec{E}$ is the electric field produced by the trap electrodes. This approximation is valid when the time scale for the variation of the ion's position and velocity is much greater than the rf period \cite{Champenois:09}.  

We define the variable $y$ as $y = z-z_0$, where $z_0$ is the chosen height of the trap center above the electrode plane. The trap center then is located at the point ($r=0$, $y=0$). The potential may be expanded in a power series about the trap center as 

\begin{multline}
\Phi(z,r) = \left( c_{y0r0} + c_{y1r0} y + c_{y2r0} y^2 + \cdots \right ) + r^2 \left(c_{y0r2} +c_{y1r2} y + c_{y2r2} y^2 + \cdots \right ) \\ + r^4 \left( c_{y0r4} + c_{y1r4}y + c_{y2r4} y^2 + \cdots \right ) + \cdots .
\label{eq:phiexp}
\end{multline}

\noindent Our goal is to construct a pseudopotential, the lowest order of which in $r$ and $y$ scales as $r^{\rho}$ and $y^{\zeta}$ for some $\rho, \zeta > 2$. By \eq{pseudopot}, $\rho$ and $\zeta$ must be even. This is tantamount to setting the terms $c_{y2r0}$ and $c_{y0r2}$ to zero. We do not venture to also null the coefficients on the ``cross terms'' which involve products of powers of $r$ and $y$. 

Understanding the relationship between $r$ and $y$ is important because $\Phi$ has no closed-form solution, except when $r=0$ \cite{THKim:10}. Let us see how we can use derivatives along $y$ to null the potential to some order along $r$. We take successively the gradient and then the divergence of \eq{phiexp}, evaluating it at the trap center. Noting that the Laplace condition states $\nabla^2 \Phi = 0$ for all $r$ and $y$, and that any term containing $r$ or $y$ to any power goes to zero at the trap center, one arrives at the following relationships: 

\begin{eqnarray}
0 = 4c_{y0r2} + 2c_{y2r0},  \nonumber \\
0 = 64c_{y0r4} + 24c_{y4r0} ...   
\label{eq:crz}
\end{eqnarray}

\noindent and so forth. We note that the form of the divergence in cylindrical coordinates implies that $c_{y0ri} < c_{yir0}$ for any $i$. These conditions assure that the cross-sections along $r$ and $y$ are super-quadratic, and that the lowest-order nonvanishing term in the expansion of $\Phi$ is at least of cubic order (in $r$ or $y$ individually, or in a product of coordinates). We aim therefore to find solutions to the set of equations of the form 

\begin{equation}
\left | \frac{\partial^k \Phi}{\partial z^k} \right |_{z = z_0} = 0, 
\label{eq:kdervs}
\end{equation}

\noindent where $k$ ranges from 1 to the desired order $n$. The $k=1$ condition implies that the electric field is zero at the trap center. As an example, if we wish to form the lowest-order multipole trap, we must null the $k=1$ and $k=2$ derivatives. To null the pseudopotential along both directions to a given order, all derivatives must go to zero. However, to null only along $r$, for instance if one is satisfied with a quartic trap along $y$, only the even derivatives greater than $k=1$ must be zero. 

\section{Constructing multipole traps}\label{sec:solving}

The equations of \eq{kdervs} are satisfied by a correct combination of the electrode widths $r_i$ or the voltages $V_i$. According to \eq{phi}, the potential is linear in each applied rf voltage; however, it is quite nonlinear in $y$ (or $z$) and in the $r_i$. Therefore, we first solve the system for the $V_i$ after fixing the electrode widths. We shall then examine the alternative approach. 

\subsection{Varying the voltages}\label{sec:voltsvary}

As a first example, let us construct a lowest-order multipole trap, meaning that the quadratic terms in $r$ and $y$, but not necessarily the cross-terms depending on products of powers of $r$ and $y$, go to zero. In addition to setting the first and second derivatives of $\Psi$ with respect to $y$ to zero at $y=0$ ($z=z_0$), we are free to choose the value of the potential at some point; therefore we specify $V_1$, the voltage of the center circular electrode: $\Phi(z=0) = V_1$. In all, we seek to satisfy the following conditions: 

\begin{eqnarray}
\left | \Phi \right |_{y = -z_0} = V_1 \nonumber \\ 
\left | \frac{\partial \Phi}{  \partial y } \right |_{y=0}=0 \nonumber \\ 
\left | \frac{\partial^2 \Phi}{\partial y^2} \right |_{y=_0} = 0 .
\label{eq:quartic_system}
\end{eqnarray}

We first choose a simple layout in which each electrode has the same width, equal to the radius of the center electrode, chosen to be 1. We also choose $V_1 = 1$ and $z_0 = 1$. Satisfying \eq{quartic_system} requires solving for two unknowns: the voltages $V_2$ and $V_3$. Taking the derivatives of $\Phi$ and forming the system, we arrive at the solution $V_2 = -0.29$, $V_3 = 4.82$. We observe that indeed the lowest-order nonvanishing term in both $\Phi(y)$ and $\Psi(y)$ is proportional to $y^4$.

\begin{figure}
\begin{center}
\includegraphics[width=.48\textwidth]{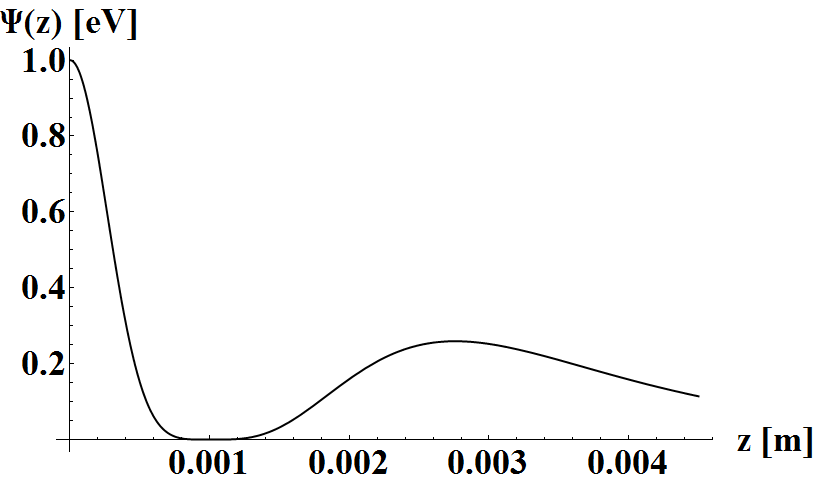}
\includegraphics[width=.48\textwidth]{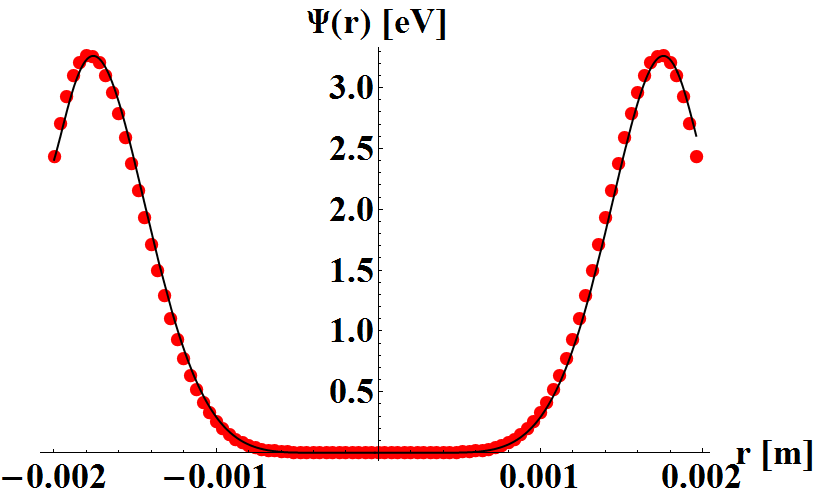}
\end{center}
\caption{Plots of $\Psi(z)$ at $r = 0$ (top) and $\Psi(r)$ at $z = z_0$ (bottom). These plots are scaled using experimental parameters given in \secref{properties}. In particular, $V_1 = 100$~V and the drive frequency $\Omega/(2\pi) = 2.86$~MHz.  Although $\Psi(r)$ is well approximated for $r \ll 0.001$~m by a function proportional to $r^6$, this fit which contains the edges of the trap contains terms up to $r^{14}$. \label{fig:psiloworder}}
\end{figure}

Our next step is to calculate the pseudopotential along $r$, which cannot be solved analytically. However, for small $r$, $\Psi(r)$ is well approximated by a polynomial containing one or two terms. Therefore, we calculate a set of points using numerical integration, and fit to a function containing those terms. In Table I, we provide the solution for traps of increasing order in $k$, and in \fig{psiloworder}, we plot $\Psi(z)$ and $\Psi(s)$ for the lowest-order multipole trap. 

\begin{table}
\begin{tabular*}{.45\textwidth}{@{\extracolsep{\fill}} | l | l | l | l | l | l | l | l |}
\hline
$k_{\mathrm {min}}$ & $\rho$ & $\zeta$ & $V_2$ & $V_3$ & $V_4$ & $V_5$ & $V_6$ \\ 
\hline
3 & 6 & 4 & -0.29 & 4.82 & & & \\
\hline
4 & 6 & 6 & 1.29 & -2.34 & 9.42 & & \\
\hline
5 & 10 & 8 & 0.97 & 2.11 & -7.60 & 15.90 & \\
\hline
6 & 10 & 10 & 1.00 & 0.77 & 5.40 & -20.30 & 28.06 \\
\hline
\end{tabular*}
\caption{Table of solutions to the system of \eq{kdervs}, with the potential nulled along both $y$ and $r$.  $k_{\mathrm {min}}$ represents the lowest order term in the expansion of $\Phi(y)$ that is nonzero. The lowest-order nonvanishing terms in $r$ and $y$ are given by $\sigma$ and $\zeta$, respectively. The number of voltages needed to satisfy the increasing number of conditions increases linearly with $k_{min}$. These solutions are only valid for the case in which each electrode has unit width (and the center electrode has unit radius), $V_1 = 1$, and $z_0 = 1$. \label{tab:table1}}
\end{table}

The voltages in \tab{table1} increase quite rapidly with increasing order $k$. Moreover, one may not be interested in nulling the potential along $z$, but only along $r$, for instance if one wishes to study planar ion crystals for which the confinement along $z$ is much tighter than along $r$. We now present solutions for the correct voltages with only \emph{even} orders nulled. Since the coefficient  $c_{y3r0}$ is not zero, we expect $\Psi$ to have a leading order in $y$ of $\zeta = 4$, even though by nulling the even derivatives, we can increase the leading order $\rho$ in $r$ to an arbitrarily high number. \tab{table2} provides a list of these solutions. We refer to the lowest-order trap described in Tables I and II as  Trap~A. 

\begin{table}
\begin{tabular*}{.45\textwidth}{@{\extracolsep{\fill}} | l | l | l | l | l | l | l | l |}
\hline
$k_{\mathrm {min}}$ & $\rho$ & $\zeta$ & $V_2$ & $V_3$ & $V_4$ & $V_5$ & $V_6$ \\ 
\hline
4 & 6 & 4 & -0.29 & 4.82 & & & \\
\hline
6 & 10 & 4 & 0.95 & -0.82 & 7.41 & & \\
\hline
8 & 14 & 4 & 0.97 & 2.17 & -7.87 & 16.19 & \\
\hline
10 & 18 & 4 & 1.00 & 1.00 & 3.92 & -17.54 & 26.56 \\
\hline
\end{tabular*}
\caption{Table of solutions to the system of \eq{kdervs}, with the potential deliberately nulled only along $r$. Here, $k_{\mathrm {min}}$ represents the lowest order nonvanishing \emph{even} term in the expansion of $\Phi(y)$. The lowest-order term altogether has $k=3$. The other symbols and parameters are as defined in \tab{table1}. \label{tab:table2}}
\end{table}

There are numerous combinations of electrode widths, $z_0$ values, and electrode voltages that will result in a multipole trap of lowest order. For example, it is possible, at least for Trap~A, to vary $z_0$ so that one of the applied voltages approaches zero. From an experimental point of view, it may be greatly preferred to limit the number of different rf voltages applied. For the same trap geometry as Trap~A, but with an ion height of 0.88, the solution is $V_2 = 1.2\times 10^{-4}$, $V_3 = 4.76$. Also, there is no reason that all electrodes must be of the same width. Indeed, in the next section we shall see that two electrodes of unequal width can carry the same rf voltage, also resulting in a multipole trap. Let us consider, as another example, a situation in which $r_1 = 1$, $r_2 = 3$, and $r_3 = 5$. For $z_0 = 1.7265$ and $V_1 = 1$, a value of $V_3 = 2.61787$ results in $V_2 = 0$.  

\subsection{Varying electrode widths}
\label{sec:varyw}

Although the most direct and convenient route to a solution is to fix the electrode widths and permit the voltages to vary, from an experimental point of view it may be  preferred to build the correct field curvature into the electrode structure, so that a single rf voltage (or ground) may be applied to all electrodes. As noted above, this problem is more difficult than solving a linear system for the unknown voltages, but we demonstrate here that it is possible, at least for a small number of electrodes. 

Let us construct a trap with $\Phi(y) \propto y^3$ in the lowest order. We assign, again, a unit width to the inner electrode ($r_1 = 1$). We set the potential of the first concentric ring to $V_2=0$, guessing that this is possible based on the result for Trap~B above. The potential of the outer ring is $V_3 = V_1 = 1$. Since we are still free to choose $z_0$, there are an infinite number of pairs ($r_2$,$r_3$) that will work. One possible solution is ($r_2 = 3.49$, $r_3 = 8.63$), which gives $z_0 = 2.46$. Again, we observe from fits to the pseudopotential that $\rho = 6$ and $\zeta = 4$. Henceforth, we refer to this trap as Trap~B. Although this approach does appear to work, we do not venture to null higher orders. 

Although, from an experimental point of view, it is inconvenient to use multiple rf signals to drive a trap, the trap fabrication will never be perfect, and there will be deviations from the solution found here due to the gapless and infinite-plane approximations \cite{Schmied:10}. Therefore, we wish to demonstrate that adjustment of the rf voltages can compensate for imperfections in the electrode geometry. We do this using a trial-and-error approach simply as a proof of concept. Let us set $r_3$ to 8.8, rather than 8.63, and vary $V_3$. We observe that $V_3 \approx 0.97$ leads again to the desired trap curvature. 

\subsection{A higher-order trap}\label{sec:highord}

We provide one further example. Higher-order traps were derived in \secref{voltsvary}, but here we again ``guess'' that the electrode widths should continue to increase further from the center in order to limit the increase in the applied voltages with increasing $r$. We then solve for the voltages. The set of radii used is ${1,3,6,10}$. Again desiring one voltage to go to zero, we set the ion height to be 2.69625, yielding a solution $V_2 = 1.28$, $V_3 \approx 0$, $V_4 = 4.50$. We refer to this as Trap~C. As a guide for the rest of the paper, we provide in \tab{examples} a summary of the three example traps. 

\begin{table}
\begin{tabular*}{.75\textwidth}{@{\extracolsep{\fill}} | l | l | l | l | l | l | l | l | l | l |}
\hline
Trap & $z_0$ & $r_1$ & $r_2$ & $r_3$ & $r_4$ & $V_1$ & $V_2$ & $V_3$ & $V_4$  \\ 
\hline
A & 1 & 1 & 2 & 3 & & 1 & -0.29 & 4.82 & \\
\hline
B & 2.46 & 1 & 3.49 & 8.63 & & 1 & 0 & 1  & \\
\hline
C & 2.69625 & 1 & 3 & 6 & 10 & 1 & 1.28 & 0 & 4.50 \\ 
\hline
\end{tabular*}
\caption{This table summarizes the five example traps that were obtained in \secref{solving}. All traps are lowest-order multipole traps except for Trap~C.  \label{tab:examples}}
\end{table}

We conclude this section by noting that the optimization of electrode shapes for construction of a multipole trap may be done in a more sophisticated way using the approach of Schmied \emph{et al.} \cite{Schmied:09}. For the lowest-order trap, the results of such calculations agree with ours \footnote{Roman Schmied, SurfacePattern software package, \url{http://atom.physik.unibas.ch/people/romanschmied.php} }.  Finally, since we have seen that imperfections in the trap dimensions can be compensated by adjustments in voltage, it may be possible to devise a scheme which varies electrode widths \emph{and} voltages, perhaps optimizing for trap depth, as well as curvature. The complication is that one may not necessarily want simply to optimize for trap depth; nulling of micromotion may be more important. The latter depends not only on the electrode widths and voltages, but also on the number and types of ions, and the drive frequency. The next section is devoted to these considerations. In particular, we perform an approximate optimization of the lowest-order trap in \secref{twoions}. 

\section{Properties of the traps}\label{sec:properties}

So far, we have shown a number of ways in which surface-electrode multipole traps can be constructed. We now consider the properties of these traps given realistic experimental parameters.  Of interest are the relationships between the overall scale of the $V_i$, the rf drive frequency $\Omega$, the trap depth $D$, and the micromotion amplitude of a particular ion, $A$. Because these traps are highly (and deliberately) anharmonic, we do not consider the ``secular frequencies'' of the ions, as these depend, in a multipole trap, on the kinetic energy (or temperature) of the ions and on the number of ions trapped. The angular modes of a circular Coulomb cluster, which can be formed in SEMT's, are described in Ref.~\cite{Lupinski:09}. 

In a quadrupole trap, the stability parameters (commonly written ``\emph{q}'' and ``\emph{a}'') are constant as long as the ion is within the quadratic region of the trap. Since the subject of this article is ion traps that require no dc voltages for confinement, we focus here on the parameter $q$. Although there are numerous regions in which ions have stable trajectories, the trap is typically operated with a $q \leq 0.3$. By contrast, with all multipole traps, one cannot simply solve for a value of $\Omega$ based on fixed formulas for trap stability. Instead, for each value of $\Omega$, one must first compute the positions of the ions in the trap. The reason is that the parameter that quantifies the stability of the ions' trajectories depends on the positions of the ions. For general multipole traps, a quantity $\eta$ which quantifies local adiabaticity is defined as 

\begin{equation}
\eta = \frac{2Q|\nabla E|}{m \Omega^2},
\label{eq:eta}
\end{equation}

\noindent where $E$ is the magnitude of the (local) electric field  \cite{Champenois:09}. The values of $\eta$ generally considered to be stable lie between 0 and 0.34, in close analogy to the parameter $q$. The micromotion amplitude $A$ for a particular ion is given by

\begin{equation}
A = \left | \frac{Q E \left (r_0 \right )}{m \Omega^2} \right |
\label{eq:micromot} 
\end{equation}

\noindent where $E\left (r_0 \right )$ is the magnitude of the electric field oriented along $\hat{r}$ at the ion location \cite{Champenois:09}. We approximate that the magnitude of the field along $\hat{z}$ is negligible. This is because, for small crystals, the ions will lie in a plane parallel to the trap electrodes, due to the stronger confinement along $\hat{z}$.  

A very simple, analytical analysis of ion behavior close to the trap center is possible with SEMT's. The reason is that, with only rf voltages required for stable ion confinement, the pseudopotential at ($y=0$) expanded about $r=0$ may be approximated by a single term. In the case of the lowest-order traps, that term is proportional to $r^6$. We wish to incorporate the length scale of the trap, represented by the ion height $z_0$, into the calculation. The potential $\Phi$ also depends linearly on the voltage scale, which we allow to be set by $V_1$. Let us then use this expression: 

\begin{equation}
\Phi(r) = \frac{c_{r} V}{z_0^4} r^4. 
\label{eq:phi4}
\end{equation}

\noindent Here, $c_{r}$ is a dimensionless constant that contains information about the effects of all the electrodes on the trap curvature, as well as the ratios of their voltages. Depending on the parameter that defines the voltage scale (for example $V_1$, or the maximal voltage on any electrode) and on how one wishes to define the length scale (for instance, using $r_1$ rather than $z_0$), the meaning and value of $c_r$ will change. 

It is now possible to find simple expressions for the other pertinent quantities, including the electric field, its derivative, and the pseudopotential. The pseudopotential $\Psi(r)$ at $y=0$ is given by 

\begin{equation}
\Psi(r) = \frac{4 c_{r}^2 Q^2 V^2}{m \Omega^2 z_0^8} r^6,  
\label{eq:psi4}
\end{equation}

\noindent and the pseudo-force $\mathcal{F}(r) \equiv - d\Psi / dr$ given by 

\begin{equation}
\mathcal{F}(r) = \frac{24 c_{r}^2 Q^2 V^2}{m \Omega^2 z_0^8} r^5. 
\label{eq:f4}
\end{equation}

We will use the quantity $\mathcal{F}$ when discussing multiple confined ions. The parameter $\eta$ at some value of $r$ may be easily computed from \eq{eta}, while the electric field may be gotten from differentiating \eq{phi4}, and $A$ obtained from that result and from \eq{micromot}. 

This basic theory can be applied readily to two simple, but important cases: a single trapped ion with some nonzero total mechanical energy, and two ions at absolute zero. In the remainder of this section, we consider these two cases, as well as crystals consisting of small numbers of ions. We assume the crystallized ions have well-defined positions, despite the cylindrical symmetry of the trap, as was observed in \cite{THKim:10}. These simple analyses shed light on some of the special properties of multipole traps, and especially SEMT's, which lend themselves easily to a simple description.

\subsection{A single ion}
\label{sec:singleion}


The region of stability for a single ion in a quadrupole trap is strongly limited by the stability parameter $q$. The basic scaling laws are as follows. While $q \propto Q V / 4 m r_0^2 \Omega^2$, the trap depth $D$ goes as $D \propto qV$. Here, $V$ is the single applied rf voltage and $r_0$ is a constant with dimensions of length that sets the scale of the trap; it may also correspond to a specific dimension (such as the distance of the ion from a particular trap electrode). Often, one sets $V$ to the maximum value attainable in one's experiment (to maximize trap depth) and then selects an $\Omega$ so that the resulting trap is stable. This effectively sets an upper limit on the ratio $Q/m$ that can be confined in the trap. 

The situation in an ideal multipole trap is very different. The parameter $\eta$  increases with distance from the center of the trap and is actually zero at the trap center. For a single ion, then, the upper limit on $Q/m$ is determined by its distance from the trap center, due to its total energy or (at absolute zero) stray fields. We can derive a set of scaling laws that describe the properties of a trapped ion. Suppose that the ion has a total mechanical energy $\mathcal{E}$. Then its maximum extent can be obtained by simply setting $\mathcal{E} = \Psi(r_{max})$: 
 
\begin{equation}
r_{max} = \left( \frac{\mathcal{E}m \Omega^2 z_0^8}{4 c_r^2 Q^2 V^2} \right) ^{1/6}. 
\label{eq:rmax}
\end{equation}

Invoking \eq{eta}, together with \eq{phi4}, we obtain an expression for $\eta$: 

\begin{equation}
\eta = 24 \left( \frac{Q c_r V \mathcal{E}}{4 m^2 \Omega^4 z_0^4} \right)^{1/3}. 
\label{eq:eta_oneion}
\end{equation}
 
\noindent As an example, let us consider some concrete numbers. The value of $c_r$ for Trap~A is 0.038. Let us assume values of $z_0 = 1$~mm, $\Omega = 2\pi \times 10^{6}$~s$^{-1}$, and $\mathcal{E} = k_B \times 10^{-3}$~J (for an ion laser-cooled to around 1~mK). For a mass of 100~amu, the charge that could be held on the ion, for $\eta = 0.3$, is $Q = 4.0 \times 10^4 e_c$, where $e_c$ is the elementary charge. It should be pointed out that an ion with any nonzero value of $Q$ less than this will also be confined, although the trap depth will fall as $Q^2$. It is not difficult to calculate $\eta$ for a single ion with any trap order $n$. It is 

\begin{equation}
\eta = \frac{2Q n^2 c V}{m \Omega^2 z_0^n} \left( \frac{4 m \Omega^2 z_0^{2n} \mathcal{E}}{Q^2 n^2 c^2 V^2} \right)^{\frac{n-2}{2n-2}}. 
\end{equation}

It is also possible to compute a value for $A$ for the single ion with nonzero energy. Using \eq{micromot} with a little algebra, we find

\begin{equation}
A = \frac{2}{\Omega}\sqrt{\frac{\mathcal{E}}{m}}. 
\end{equation}

\noindent This expression does not depend on the charge $Q$. Therefore, there is no disadvantage from the standpoint of micromotion to storing a highly-charged ion. This expression also holds for all orders of traps, while e.g. \eq{eta_oneion} holds only for the lowest-order traps. Remarkably, $A$ also does not depend on  details of the electric potential such as $c_r$ and $V_1$. The implication is that the lowest-order multipole trap is as good as a higher-order trap, from the standpoint of micromotion reduction, but only for a single ion. 

\subsection{Two cold ions}
\label{sec:twoions}

Another case that lends itself easily to analysis is the confinement of two ions at absolute zero. The ions, by Coulomb repulsion, are each located some distance $r_{eq}$ from $r=0$. We assume, again, that the two ions lie in the plane parallel to the trap electrodes. This is approximately true in a real trap. We begin with some examples designed to elucidate the unique properties of SEMT's, and then present the general theory. 

\begin{figure}
\begin{center}
\includegraphics[width=.65\textwidth]{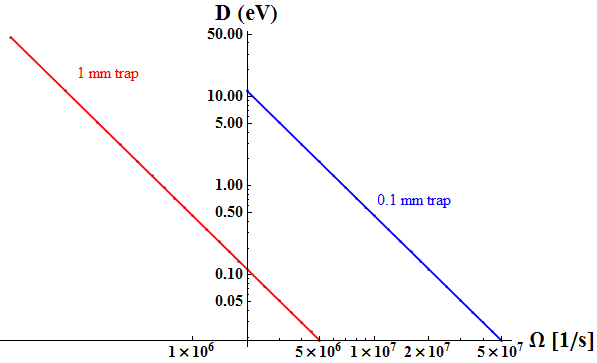} \\ 
\includegraphics[width=.65\textwidth]{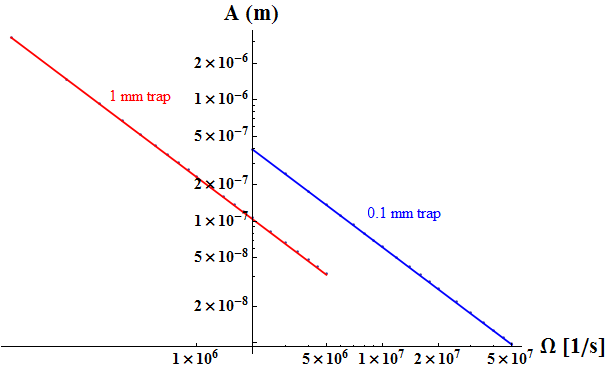}
\end{center} 
\caption{Logarithmic plots of the trap depth $D$ (left) and micromotion amplitude $A$ (right), as functions of $\Omega$, for two $^{88}\mathrm{Sr}^+$ ions confined in Trap~A with $z_0 = r_1 = 1$~mm and $V_1 = 100$~V. The trap depth scales as $D \propto \Omega^{-2}$, and the amplitudes as $A \propto \Omega^{-1.15}$. 
\label{fig:DA_scaling}}
\end{figure} 

We first compute values of $\Omega$, $A$, and $D$ for Trap~A. Rather than an exhaustive search over all possible parameters, we examine a few special cases. In particular, we consider two length scales: $r_1 = 1$~mm and $r_1 = 0.1$~mm. We choose a voltage scale $V_1 = 100$~V, resulting in $V_2 = -29$~V and $V_3 = 482$~V. Finally, we assign to the ion the charge and mass values of $^{88}\mathrm{Sr}^+$. Plots of $D$ and $A$ as functions of $\Omega$ are given in \fig{DA_scaling}. Unsurprisingly, the trap depths follow a simple proportionality to $1/\Omega^2$, according to \eq{pseudopot}. The amplitudes follow a different power law, roughly $\Omega^{-1.15}$, which is explained below. It is most important to note that the highest depth corresponds to the highest level of micromotion, and vice-versa. We can also identify a reason why one might wish to scale these traps down. With $z_0 = 1$~mm, an appropriate drive frequency is 1.5~MHz, leading to $A = 148$~nm. With $z_0 = 0.1$~mm, by contrast, a drive frequency of 15~MHz (which leads to the same trap depth) gives $A = 39$~nm. It is useful to compare this to the situation in a quadrupole trap, in which $A \approx q r_{eq}/2$, where $r_{eq}$ is the distance from the rf null, and $q \approx 0.3$ is a typical stability parameter. Let us introduce a relative micromotion amplitude $\aleph = A/r_{eq}$. The idea is to quantify the level of micromotion, taking into account the overall size of the ion crystal. For quadrupole traps, then, $\aleph \approx 0.15$ (in the radial direction, in the case of a linear trap) under typical conditions. For two ions in a quadrupole trap, a typical value of the distance of each from the trap center is $r = 2$~$\mu$m; in this case, $A$ = 300~nm. Let us choose a small, but experimentally feasible value for the trap depth: $D = 100$~meV. The frequency that leads to this depth, for Trap~A, is 4.6~MHz, yielding $r_{eq} = 129$~$\mu$m and $A = 58$~nm, for a result of $\aleph = 4.5\times 10^{-4}$. The value of $A$ relative to the crystal size is orders of magnitude smaller than in a quadrupole trap, meaning that \emph{interior} ions in a multi-ion crystal should have correspondingly suppressed values of $A$. 

One remarkable feature of the data is the very high trap depth possible in a surface-electrode multipole trap. As an example, the depth of the trap used in Ref.~\cite{THKim:10} is approximately 170~meV, which is typical for millimeter-scale surface-electrode ion traps. For Trap~A with $r_1 = 1$~mm, we predict a depth of 50~eV when the trap is driven at 100~kHz with rf voltages of $V_1 = 100$~V, $V_2 = -13.66$~V, and $V_3 = 225.23$~V. The reason why such a deep trap can be made is that the stability parameter $\eta$ is not a constant within the trapping region, as is the case for quadrupole traps (in the region within which the electric potential is approximately quadratic), but depends in a nonlinear manner on $r$. 

Other example traps may be analyzed in a similar manner. Because of the somewhat large parameter space, we focus our attention. We choose a length scale of 1~mm and set the depth to 100~meV. We also select a maximum electrode voltage, choosing a common laboratory value of 300~V. The values of $r_{eq}$, $D$, and $\eta$ for one of two ions in each example trap are given in \tab{example_compare}. We do not further comment on the optimization of the ion height for a particular trap geometry; simple numerical experiments show that, at least for Trap~A, $z_0$ is already near the optimal value, given fixed values of the electrode radii. In this paper, we do not endeavor to present a truly ``optimal'' trap, since the \emph{criteria} for what makes a trap optimal depend on the goals of a particular experiment. It may be possible, even just for the two-ion case, to improve on $A$ for a fixed trap depth by searching the entire parameter space of $r_2$, $r_3$, and $z_0$. Once these are fixed, $V_2$ and $V_3$ are uniquely determined, given an overall voltage scale $V_1$. 

\begin{table}
\begin{center}
\begin{tabular}{| l | l | l | l | l | l |}
\hline 
Trap & $\Omega/(2\pi)$ [MHz] & $A$ [nm] & $\eta$ & $r_{eq}$ [$\mu$m] & $\aleph$ \\
\hline
A & 4.6 & 58 & $1.0 \times 10^{-4}$ & 129 & $7.5\times 10^{-4}$ \\ 
\hline
B & 2.55 & 121 & $1.9 \times 10^{-4}$ & 194 & $6.2\times 10^{-4}$ \\
\hline
C & 1.24 & 919 & $8.3\times 10^{-3}$ & 213 & $4.3\times 10^{-3}$ \\ 
\hline
\end{tabular}
\end{center}
\caption{Table of values for the example traps. In all cases, the \emph{maximal} rf voltage on any electrode is 300~V, the length scale is set by $r_1 = 1$~mm, and the trap depth is 100~meV.  \label{tab:example_compare}}
\end{table}

\begin{figure}
\begin{center}
\includegraphics[width=.8\textwidth]{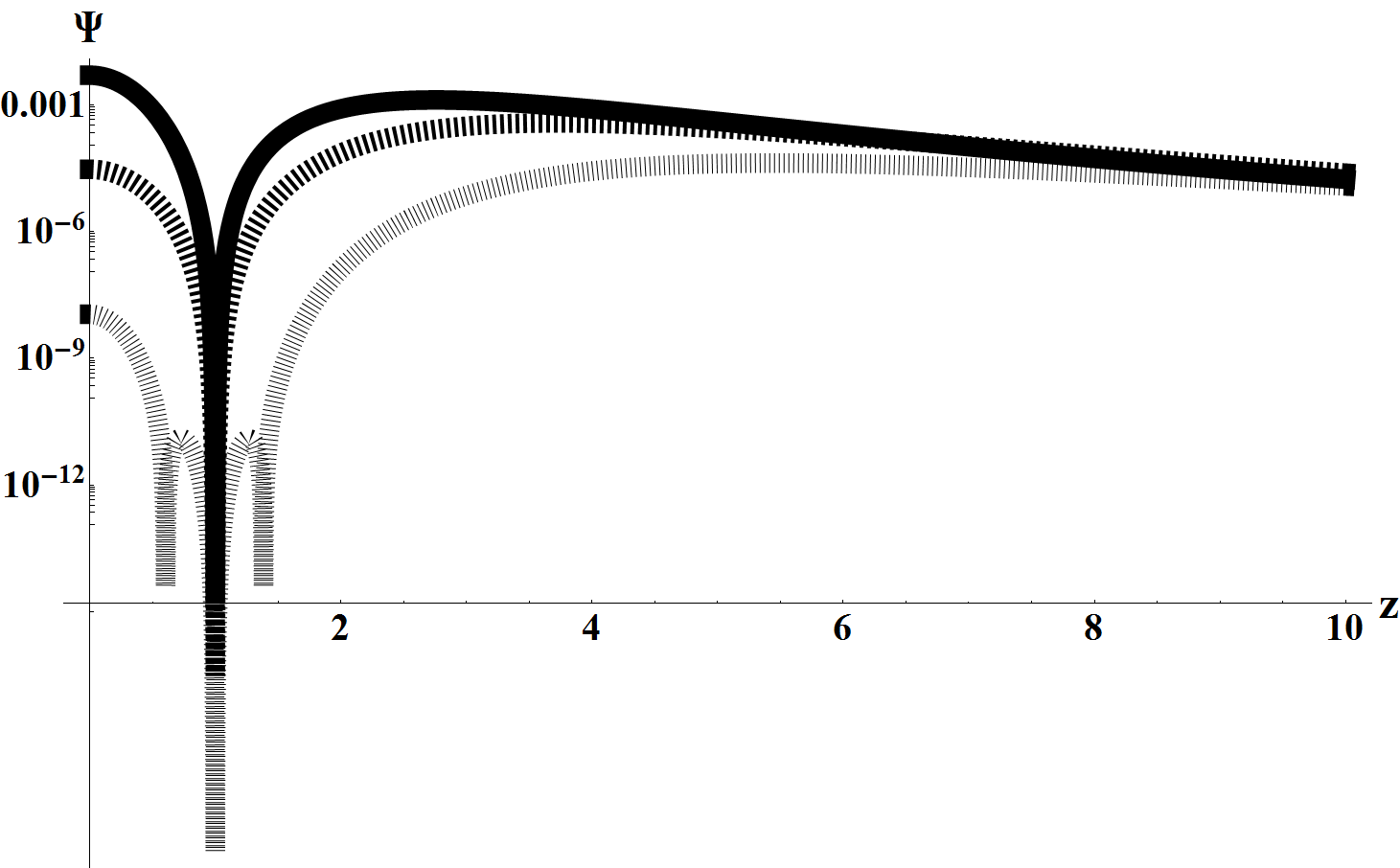}
\end{center} 
\caption{Logarithmic plots of $\Psi(z)$ for Trap~A for successive possible orders of $\rho$. The solid line represents the $\rho=6$ trap, thick-dashed represents $\rho=10$, and thin-dashed represents $\rho=14$. The length scale for all traps is $r_1=1$ with $z_0=1$. The voltages for each trap are normalized such that the highest rf voltage applied to any trap electrode is 1. Units are arbitrary. 
\label{fig:depth_comparison}}
\end{figure}

We now consider higher-order traps. We observe that although the dependence of the pseudopotential on $r$ and $y$ is as expected, the trap depth falls precipitously as the order is increased. A plot illustrating this is given as \fig{depth_comparison}. Strictly speaking, the traps should first be optimized in some way, but based on numerous numerical experiments (including different methods of optimization), we think it likely that the unscaled depth is much smaller for all higher-order traps. The trap depth seen in \fig{depth_comparison} is not absolute; the real value depends on experimental parameters. Suppose we drive the trap at the highest voltage achievable, and then reduce $\Omega$ until the trap depth is some minimum acceptable value. $A$ increases with $\Omega$ just as $D$ does. For Trap~C, listed in \tab{example_compare}, $A$ is roughly an order of magnitude higher than for a corresponding lower-order trap with the same trap depth and maximal voltage.  $\aleph$ is also roughly an order of magnitude higher than the best lowest-order traps. Therefore we conjecture, that for all practical purposes, lowest-order multipole traps are actually the best choice. Given sufficient optimization of the higher-order traps, however, this could turn out not to be true.  

Returning to the lowest-order trap, we again assume the trap potential is given by \eq{phi4}. To obtain the general formulas for $\eta$ and $A$, we set \eq{f4} equal to the Coulomb force between the two ions. The equilibrium ion displacement $r_{eq}$ is then

\begin{equation}
r_{eq} = \left [ \frac{k_e m \Omega^2 z_0^8}{96 c_{r}^2 V^2} \right ]^{1/7},  
\end{equation}

\noindent where $k_e$ is the Coulomb constant: $k_e = 8.988\times 10^9$~N$\cdot$m$^2$/C$^2$. Taking the first and second derivatives of $\Phi(r)$ and using \eq{eta} and \eq{micromot}, and substituting in our expression for $r_{eq}$, we obtain the following expressions for $\eta$, $A$, and $\aleph$: 

\begin{equation}
\eta = 24Q \left [ \frac{ V^3 k_e^2}{96^2 c_r m^5 \Omega^{10} z_0^{12}} \right ]^{1/7}
\label{eq:eta4} 
\end{equation}

\begin{equation} 
A = 4Q \left [ \frac{c_r V k_e^3}{96^3 m^4 \Omega^{8} z_0^{4}} \right ]^{1/7}. 
\label{eq:A4} 
\end{equation}

\begin{equation}
\aleph = 4Q \left ( \frac{c_r^3 V^3 k_e^2}{96^2 m^5 \Omega^{10} z_0^{12}} \right )^{1/7}
\end{equation}

We can immediately check this model by examining the scaling of $A$ with the frequency $\Omega$. A fit to our data for Trap~A shows that $A \propto \Omega^{-1.15}$, and the predicted exponent in \eq{A4} is $-8/7 \approx -1.14$. We take this to mean that we can indeed approximate $\Psi(r) \propto r^6$ in the region of space occupied by two ions. Further confirmation has been obtained by comparing the values of $A$ and $\eta$ computed by \eq{A4} and \eq{eta4}, respectively, to the values found by numerically computing $E(r_{eq})$ and $\nabla E(r_{eq})$ according to the value of $r_{eq}$ found numerically. The values agree to within one percent, the error coming most likely from imperfect fits to the results of the numerical integration for $\Psi(r)$. 

Let us consider also the effect of the overall voltage scale: $A \propto V^{1/7}/\Omega^{8/7}$. Using $D \propto V^2/\Omega^2$, we see that one can reduce the micromotion by raising both $V$ and $\Omega$ so that their ratio remains constant. This is true in quadrupole traps as well, but with different powers in the scaling laws ($\eta \propto V/\Omega^2$ and $A \propto V^{2/3}/\Omega^{4/3}$). In practice, technical limitations will dominate the suppression of micromotion, even in the lowest-order traps. To give an example, suppose we want to use Trap~A with $z_0=0.1$~mm. With $V_1 = 623$~V and $\Omega = 2\pi \times 286$~MHz, $A$ falls to 2.6~nm, with $\eta = 8.6\times 10^{-4}$. These values, however, are already technically challenging. 

These scaling arguments highlight the marked differences between multipole and quadrupole traps. Whereas in a quadrupole trap the stability parameter $q \propto V/\Omega^2$ and is a constant over the entire trap volume, the equivalent parameter $\eta$ depends strongly on the ion position, and in our examples here is far below 0.3. Furthermore, in a quadrupole trap, the pseudo-force $\mathcal{F}$ and the magnitude of the real electric field along $r$, $E_r$,  are both linear in the coordinate $r$. By contrast, $\mathcal{F}$ always scales up as a higher power of $r$  than $E$, meaning that the ion is pinned to a significantly lower region of electric field than in a comparable quadrupole trap. This is, fundamentally, the reason for the desirable properties of ideal multipole traps: small values of $\eta$, large possible values of $D$, and/ or the suppression of $A$. 

\subsection{Multiple ions}
\label{sec:crystals}

\begin{figure}
\begin{center}
\includegraphics[width=.65\textwidth]{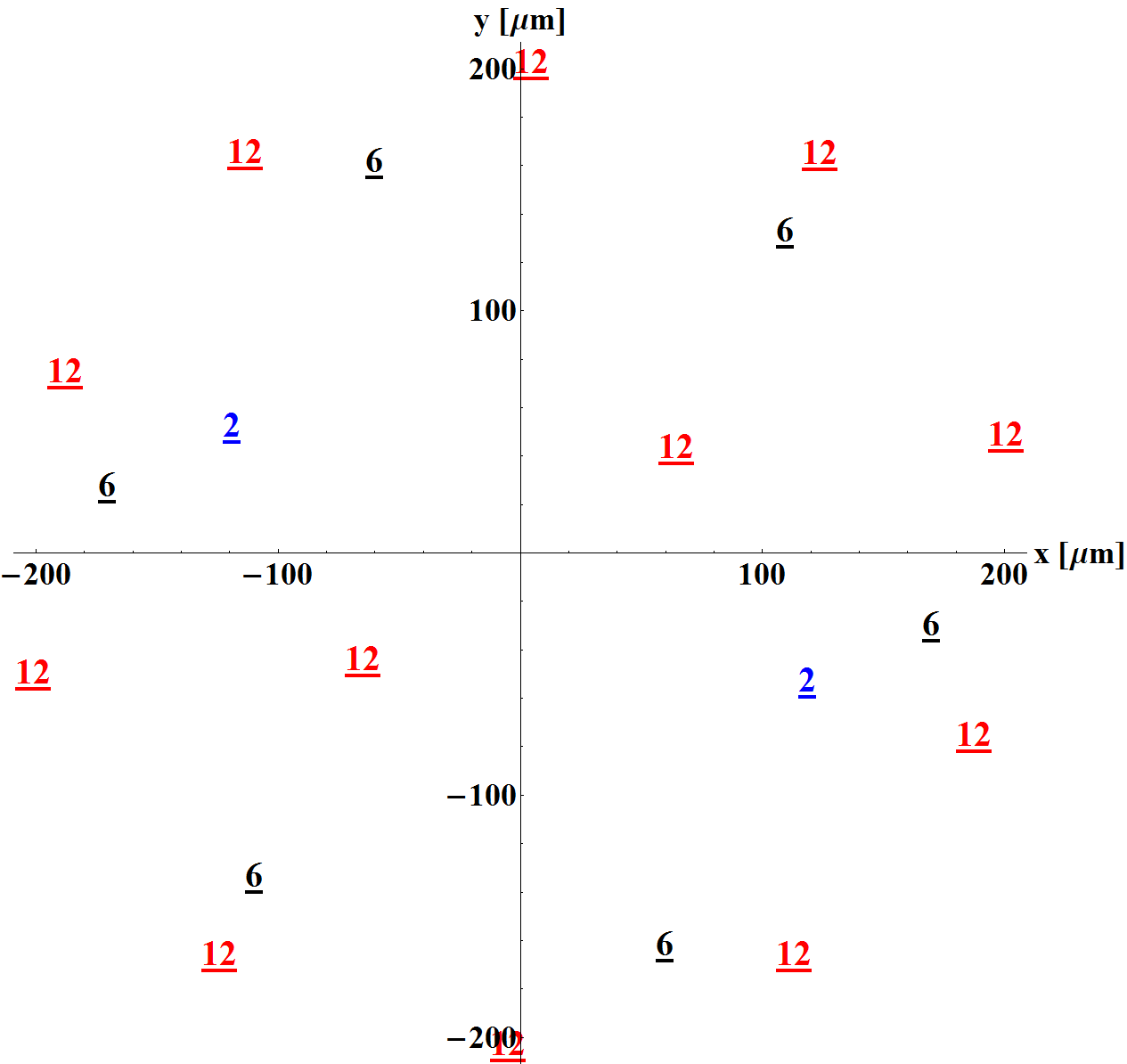} \\ 
\end{center} 
\caption{Plots of crystal structures in Trap~A with $r_1=1$~mm, $V_1 = 100$~V, and $\Omega/(2\pi) = 4.6$~MHz, leading to a trap depth of 100~meV. Crystals of two, six, and twelve ions are represented, and the ion locations in each crystal are plotted with a ``\underline{2},'' ``\underline{6},'' and ``\underline{12},'' respectively. In all cases, the displacement of each ion from $z = z_0$ is 17~$\mu$m or less, and thus the crystals are approximately planar. (This displacement increases with the number of ions.)
\label{fig:crystals}}
\end{figure}

We now turn briefly to the question of the structure of multi-ion crystals in our multipole traps, and the related issues of stability and micromotion. We assume again that the temperature of the ions is zero, and find the positions that minimize the energy of the crystal. The results, not surprisingly, resemble the structure of crystals in linear multipole traps. In this article, we consider only the structures of small numbers of ions that form a ring in the plane parallel to the trap electrodes, assuming that $\Psi(y) \propto y^4$. We also specify the radial pseudopotential using \eq{psi4}. We present in \fig{crystals} three representative crystal structures for Trap~$\mathrm{A}$. 

We may now easily calculate the micromotion amplitudes for ions in such 2-D crystals by calculating the magnitude of the electric field at the ion's location. We reference the crystals plotted in \fig{crystals}. For the two-ion crystal, $A = 61$~nm. For six ions, $A = 142$~nm, and for ions in the outer shell of the 12-ion crystal, $A=244$~nm. However, the value of $A$ for the \emph{interior} ions is $A = 13$~nm. This raises the notion of using an outer shell of ions perhaps only for sympathetic cooling, while performing precision operations only on a ring of interior ions, if errors due to micromotion are a concern. 

\section{Conclusions}\label{sec:discuss} 

We have shown, for the first time, a way to create an ideal multipole ion trap, in which ions are repelled from the rf null only by Coulomb repulsion and not by dc electric fields. Moreover, this is the first time that a surface-electrode multipole trap has been proposed. Although the basic procedure for solving for the  voltages and/or electrode dimensions that null the quadratic terms in the pseudopotential is fairly straightforward, there is a large parameter space (including electrode widths and voltages, ion height, overall length and voltage scales, and drive frequency), which can make the route to optimization quite complicated. Here, we have tried to give a summary of some of the key features of these traps. 

Many questions remain. It would be advantageous to develop an automated routine that can optimize a trap under certain criteria, for certain numbers of ions. It is also not known whether it is possible to design higher-order SEMT's that are as deep as the lower-order versions while offering superior micromotion suppression.  Additional concerns are the structure of multi-ion crystals in these traps, their motional frequencies and coupling rates, and their micromotion amplitudes. The computations of these quantities will require more sophisticated procedures and were beyond the scope of this paper. 

There are many potential applications of this work. The fact that no dc confinement is required means that micromotion, a source of systematic error (\emph{e.g.} in atomic clocks \cite{Champenois:10}), can potentially be lower than in linear multipole traps. There is also the possibility of using SEMT's for the study of cold atom-ion collisions, since currently the temperature to which the mixture can be cooled is limited by the rf-induced heating of the ions \cite{Cetina:12}. The utility of these traps for quantum computation and quantum simulation should also be considered. Since these traps are amenable to microfabrication, one could envision scaling down multipole ion traps, in a manner similar to that done with linear quadrupole ion traps in recent years. One implication is that ions can be placed near surfaces for quantum simulation using rf or microwave fields, in combination with magnetic field gradients \cite{Wunderlich:08, Ospelkaus:08}. This is in addition to the benefit discussed in \secref{properties}, which is that micromotion amplitudes are suppressed, all else being equal, when the trap is scaled down. In the end, the applications seem numerous, and this list does not attempt to be exhaustive. 

We gratefully acknowledge funding from The Citadel and The Citadel Foundation and  helpful discussions with, Roman Schmied, Caroline Champenois, David Kielpinski, and Kenneth Brown. 

\bibliographystyle{apsrev}

\end{document}